\newcommand{\CLASSINPUTtoptextmargin}{2.4cm}
\newcommand{\CLASSINPUTbottomtextmargin}{4.6cm}

\documentclass[conference,letter]{IEEEtran}
\IEEEoverridecommandlockouts
\usepackage{cite}
\usepackage{amsmath,amssymb,amsfonts}
\usepackage{algorithmic}
\usepackage{graphicx}
\usepackage{textcomp}
\usepackage{xcolor}
\usepackage{tikz}
\usepackage{graphicx}
\usepackage{pgfplots}
\usepgfplotslibrary{groupplots}
\pgfplotsset{compat=1.12}
\usepackage{subfig}
\usepackage{algorithmic}
\usepackage{algorithm}
    \makeatletter 
\newcommand{\linebreakand}{%
  \end{@IEEEauthorhalign}
  \hfill\mbox{}\par
  \mbox{}\hfill\begin{@IEEEauthorhalign}
}
\makeatother 

\begin{document}

\title{Ransomware: Analysis and Evaluation of Live Forensic
Techniques and the Impact on Linux based IoT Systems}

\author{\IEEEauthorblockN{Salko Korac, Leandros Maglaras, Naghmeh Moradpoor, Bill Buchanan, Berk Canberk}
\IEEEauthorblockA{\textit{Edinburgh Napier University}, 
\textit{Edinburgh EH10 5DT, UK }\\
(40517266, l.maglaras, n.moradpoor, b.canberk, b.buchanan)@napier.ac.uk}
}

\maketitle

\begin{abstract}
Ransomware has been predominantly a threat to Windows systems. But, Linux systems became interesting for cybercriminals and this trend is expected to continue. This endangers IoT ecosystems, whereas many IoT systems are based on Linux (e.g. cloud infrastructure and gateways). This paper researches how currently employed forensic techniques can be applied to Linux ransomware and evaluates the maturity as well as the impact on the system.  While Windows-based ransomware predominantly uses RSA and AES for key management, a variety of approaches was identified for Linux. Cybercriminals appear to be deliberately moving away from RSA and AES to make Live forensic investigations more difficult. Linux ransomware is developed for a predefined goal and does not exploit the full potential of damage. It appears in an early stage and is expected to reach a similar potential to Windows-based malware. The results generated provided an excellent basic understanding to discuss and assess implications on the IoT industry at an early stage of development. 
\end{abstract}


\section{Introduction}
The advancement of low-cost computers, cloud services, big data technologies, analytics, and mobile technologies has made it possible for small physical devices to build networks and exchange data without the need for human intervention.  IoT device interconnection exposes users to various security risks in addition to efficiency and accuracy issues when connecting to vital systems. Linux systems became interesting for cybercriminals and this trend is expected to continue. This endangers IoT ecosystems, whereas many IoT systems are based on Linux (e.g. cloud infrastructure and gateways). As stated in ENISA threat landscape reports, ransomware is one of the major threats against digital systems \cite{lella2023enisa}.

The 1989 AIDS Trojan was the first ransomware. This ransomware claimed to educate
about the autoimmune disease and was distributed via floppy disk. Before installation,
users had to agree to a license, which required a payment of few hundred US dollars \cite{ryan2021ransomware}.

Nowadays, ransomware has become a major threat with revenue of USD 765.6 million in
2021 \cite{chainalaysis}. The United States White House wants to take a financially-focused approach to curbing
the ransomware problem and has launched the International Counter Ransomware
Initiative (CRI). As of October 2023, 50 member countries, including the United
Kingdom, have pledged not to pay a ransom \cite{thewhitehouse}. This initiative
would urge member countries to take appropriate proactive measures instead of paying the ransom.

Ransomware payments became more legally risky as some of the
ransomware gangs were linked to sanctioned organisations. According to criminal
investigations by Europol, the Ukrainian-Russian war forced cybercriminal gangs to
relocate their activities into other jurisdictions \cite{europol}. There is widespread agreement that Russian cybercriminals do not target
victims in Russia. According to Glenny, the Russian cybercriminal group Cl0p avoids
attacks on government institutions, cities or the police \cite{glenny}. 


\subsubsection{\textbf{Relevance for Linux systems.}} The market share of Windows operating system for desktops decreased from 95.42\% in January
2009 to 69.52\% in July 2023 \cite{statistamarketshare}. At the same time, the Unix-based macOS \cite{thaeler2023enhancing} and Linux-based desktop systems increased the cumulative share
from 4.33 to 23.54\%. But, Linux holds a significant server market share estimated between 62.4\% and 70.4\% \cite{statistamarketshareserver, fortunebusiness}. Linux systems became more interesting for cybercriminals. Security researchers such as Terefos have observed Windows-based Cl0p ransomware being expanded to attack Linux systems \cite{terefos}. Also other
industry actors observe that Linux systems are becoming a prime target for ransomware
\cite{gdata}. And the
ransomware trend for Linux is expected to continue \cite{trendmicro}. 

\subsubsection{\textbf{Attack chain.}} The attackers improved the organisation and industrialisation of ransomware extortion
by adopting a Ransomware as a Service (RaaS) model \cite{johansen}.
Due to the increasing division of tasks and industrialisation of various work steps in the
field of cybercrime, many attacks aim to gain initial access to the victim's network and later it will be decided how to utilise the access. Ransomware should no longer be viewed as a stand-alone software product.
Increasingly, entire chains of attacks take place before the ransomware is executed.
Based on the professional experience and the above explanation, an attack chain is
introduced as shown in Figure \ref{fig:attackchain} below. 

\begin{figure*}[htp]
    \centering
    \includegraphics[width=15.84cm]{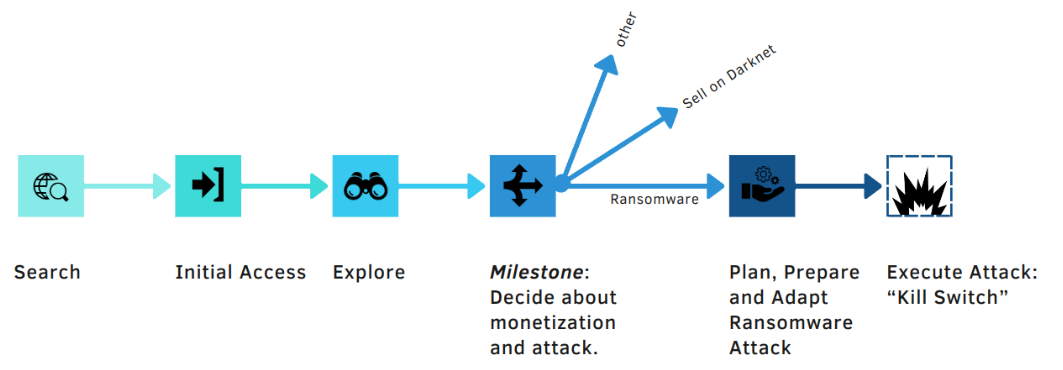}
    \caption{Attack chain with conscious decision about further use}
    \label{fig:attackchain}
\end{figure*}

This attack chain model considers a conscious
decision milestone by the hackers, where further use is discussed and decided.
Specialised cybercriminals focus on searching for new victims and gaining initial
access. This is achieved through targeted attacks using individual methods and through
non-targeted attacks using automated processes (e.g. brute force techniques, exploits,
zero-day exploits). 

\subsubsection{\textbf{The emotional aspect during a ransomware attack}}
\begin{figure*}[htp]
    \centering
    \includegraphics[width=15.84cm]{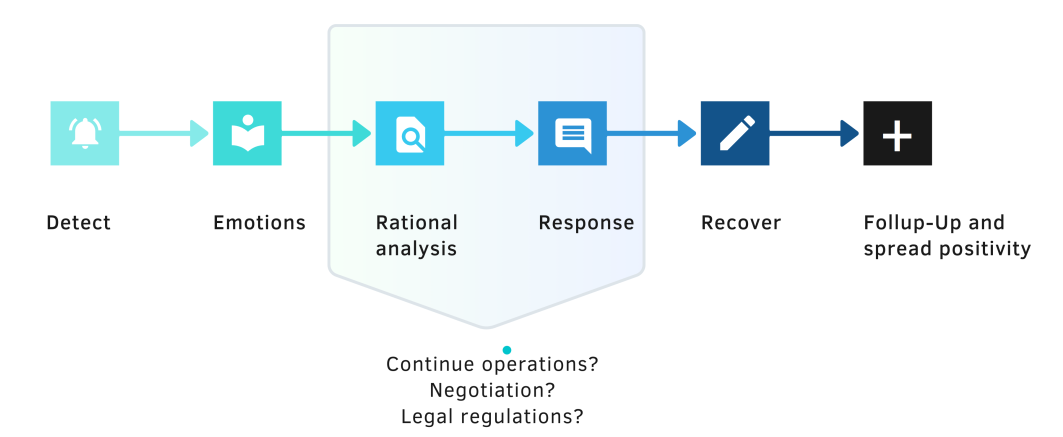}
    \caption{Response chain from victim's view}
    \label{fig:victimschain}
\end{figure*}

It is known from professional experience that ransomware victims initially react
emotionally and categorically refuse to negotiate. After taking stock, victims react more
soberly, are more likely to consider paying the ransom and are more open to recognising
this process as a deal.
Current understandings of ransomware attacks had painted a very rational picture how
the attacks happen. Little or no consideration is given to the emotional component of
both the victim and the attacker during the literature review. Figure \ref{fig:victimschain} shows the response chain from the victim's point of view.

The victim undergoes an emotional phase and moves to a rational analysis after a certain period of time. Internal power struggles and lack of cooperation lead to a significantly longer resolution time \cite{internalbsipaper}.  The influence of emotional
factors on cyber incidents would be an interesting further field of research.

The aim of this work is to examine the current maturity level of ransomware on Linux
operating systems. During the experiments, commonly used live forensic techniques
were applied and the results were compared with ransomware for Windows operating
systems. 

The techniques described in this work are intended to demonstrate techniques that can
be used to mitigate an ongoing ransomware attack on Linux operating systems. This
topic was chosen because ransomware for Windows operating systems is currently
widespread, but ransomware for Linux is expected to increase significantly.

The results allow conclusions for the protection of IoT solutions, which many are based on Linux operating systems.

\section{Experiments design}
The experiments design considered 3 dimensions: 3 ransomware sample were executed on 2 different Linux operating systems with 2 different permission levels. Overall 12 combinations, in 24 main executions plus 6 retests were covered by the testplan.

\textbf{Is the key in memory?} This experiment runs the ransomware and creates regular memory dumps for analysis using
live forensic techniques. 

A sample of the ransomware execution is shown below. \\ 
{\fontfamily{pcr}\scriptsize\selectfont
\# ubuntu@server-clean: ./blackbasta.elf
}
\\
The memory dumps were taken before and during test execution and after reboot. The memory was extracted with VirtualBox internal dump functions.
\\ 
{\fontfamily{pcr}\scriptsize\selectfont
\# VBoxManage debugvm "Client-Debian" dumpvmcore --filename \$ransomwaresample.1st.run.5seconds.elf
}
\\
Each memory dump was analyzed for the existance of the key using at least three forensic tools.
\\ 
{\fontfamily{pcr}\scriptsize\selectfont
\# aeskeyfind Server-Clean-CMD-Test11-beforeRun.elf
}
\\
{\fontfamily{pcr}\scriptsize\selectfont
\# rsakeyfind Server-Clean-CMD-Test11-beforeRun.elf
}
\\
{\fontfamily{pcr}\scriptsize\selectfont
\# findaes Server-Clean-CMD-Test11-beforeRun.elf
}

Any keys found were documented and compared before and after. The tests were repeated on the retest environment and before ransomware execution it was ensured, that the memory did not contain any AES or RSA keys in memory.

\textbf{How long is the key present.} In this experiment memory dumps were taken in 7 time intervals with the goal to identify
how long the key is present in the memory. Memory dump is also taken before the
experiment and after reboot. 
A script was developed, which generated 6 dumps after 5, 35, 65, 95 and 125 seconds as well as after 15 minutes after execution of the ransomware.

\textbf{Does the key decrypt the files.} If keys are found in memory, they are used to decrypt the files. The decryption experiment followed the work done and the script developed by Davies \cite{davies2020}

\textbf{Does the sample spread throughout the network.} This experiment examines whether and how the ransomware attempts to spread across
the network. Network communications must be captured outside the virtual machine to
prevent detection and evasion by the ransomware. To simulate a realistic attack surface,
a second server is placed on the same network providing SSH, FTP, Web services,
MySQL databases and a Samba file share. During the experiment execution, client and
server will have open connections through these protocols.

The network communication was traced on all active virtual machines with VirtualBox internal functions, as can be seen, for example, in the command below. 

{\fontfamily{pcr}\scriptsize\selectfont
\# VBoxManage modifyvm "ubuntu" --nictrace1 on 
}

{\fontfamily{pcr}\scriptsize\selectfont
--nictracefile1 out.pcap
}

{\fontfamily{pcr}\scriptsize\selectfont
\# VirtualBox -startvm "ubuntu"\\
}

\textbf{What is the impact of the encryption}. In this experiment the impact on the system will be analysed. Honeypot files will be
placed in /home, /root, /var/lib/mysql, /var/www, /etc/nginx, -/apache2 and
/var/log. Also a web server with a small web page will be created as honeypot.
Furthermore a server will provide FTP, SSH and Samba file shares and will have open
connections with a client. File hashes will reveal which files were encrypted by the ransomware.

\textbf{Test plan.} A multi-dimensional combination was developed that takes into account different
operating systems and permission levels.

\textbf{Playbook.} To ensure consistent experiment quality, a playbook was defined that is used to prepare
and conduct each experiment.

\begin{figure}[htp]
    \centering
    \includegraphics[width=8.84cm]{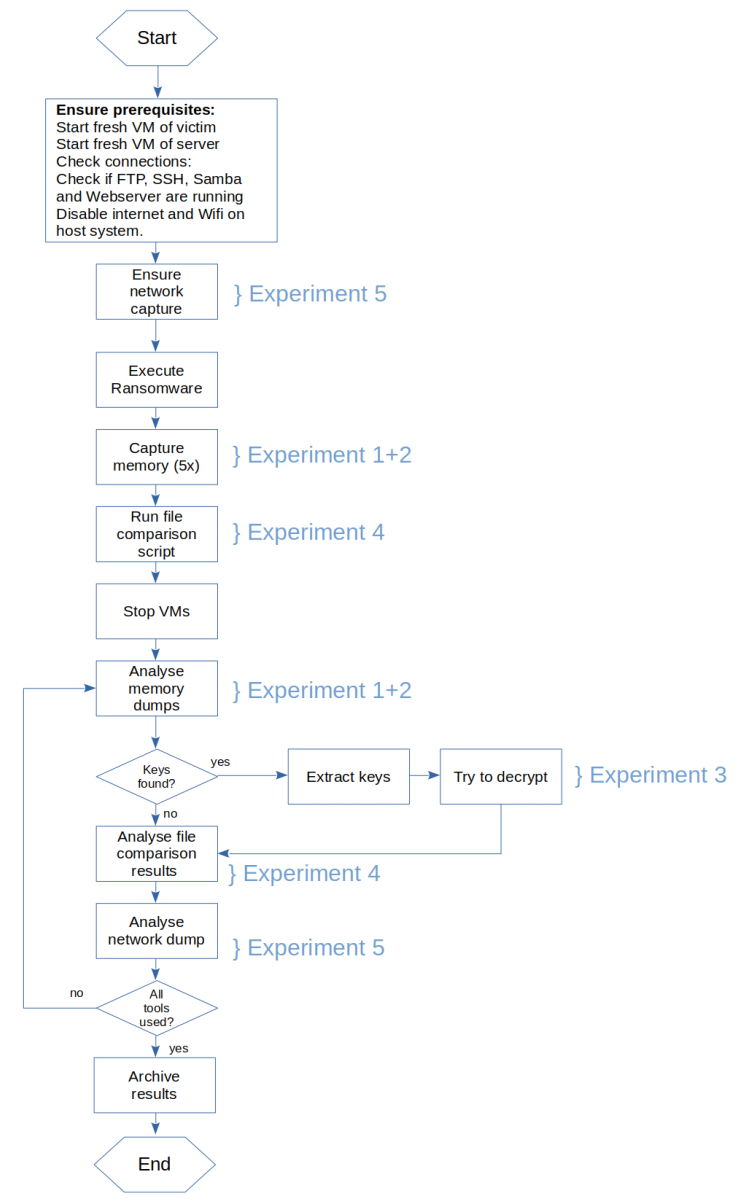}
    \caption{Playbook for experiment execution}
    \label{fig:playbook}
\end{figure}

The general steps in the playbook are shown in Figure \ref{fig:playbook}.

\section{Environment design}
When designing the test environment, care must be taken to ensure that it is realistic. The design should only deviate from real conditions if necessary. When designing a test
environment, a careful balance must always be struck between effort and the creation of
realistic conditions. Virtual machine technology was used in this project because cyber criminals have to take
into account that Linux is often operated in virtual cloud environments. This provides a
convenient way to perform external memory dumps as well as external network capture,
without noticeable risk of being discovered by the ransomware. During execution it was found out, that the initial environment design was too realistic.
Forensic investigations took significantly more time than planned. If unexpected or suspicious
results were encountered, a retest environment was required to validate the results and to eliminate disruptive factors.
The virtual hardware is displayed in Table \ref{hardware}.

\begin{center}
\begin{tabular}{ l l l}
 \textbf{Machine Name} & \textbf{Operating System}  & \textbf{Purpose}  \\
   \hline
 Server-Clean       & Ubuntu 20.04.6    & Victim \\  
 Client             & Debian 12.1.0     & Victim  \\ 
 Server-Clean-CMD   & Ubuntu 20.04.6    &  Retest
\end{tabular}
\captionof{table}{Virtual Hardware Configuration}\label{hardware}
\end{center}

It was important to protect unintended spread through the network. \\
{\fontfamily{pcr}\scriptsize\selectfont
sudo ifconfig vboxnet0 down
}.  \\ The host network interface was deactivated with above-mentioned command on the host machine.

\section{Implementation}

Several recent ransomware attacks were researched. Following Linux ransomware samples were identified:

\textbf{Icefire.} It was primarily designed and developed for Windows, but in
the recent months also attacks on Linux operating systems have been observed misusing
IBM Aspera Faspex file transfer software \cite{delamotte}.

\textbf{Cl0p.} It is known to be similar to it’s Windows variant and appears to
be still in initial development \cite{terefos}.

\textbf{Blackbasta.} This sample is linked to be a sub-variant of the Conti ransomware group and
focuses on exploiting VMware's ESXi virtual machine technology \cite{umawing}.

The hashes of all samples are listed in Table \ref{linuxsamples} below.

\begin{center}
\begin{tabular}{ l l }
 \textbf{Name} & \textbf{SHA-256 hash}  \\ 
   \hline

 Icefire    &  e9cc7fdfa3cf40ff9c3db0248a79f48 \\
            & 17b170f2660aa2b2ed6c551eae1c38e0b  \\  
 Cl0p       &     09d6dab9b70a74f61c41eaa485b37de \\
            &     9a40c86b6d2eae7413db11b4e6a8256ef  \\ 
 Blackbasta & 0d6c3de5aebbbe85939d7588150edf7 \\
            & b7bdc712fceb6a83d79e65b6f79bfc2ef  
\end{tabular}
\captionof{table}{Linux ransomware samples}\label{linuxsamples}
\end{center}

\begin{figure*}[htp]
    \centering
    \includegraphics[width=15.84cm]{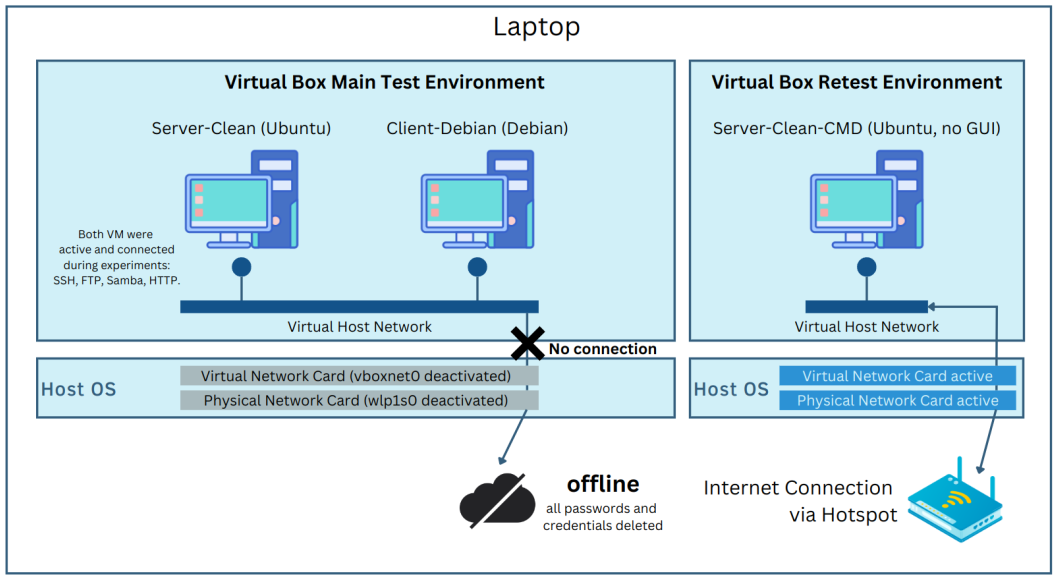}
    \caption{Environment design, including test environment}
    \label{fig:network}
\end{figure*}

The network topology is shown in Figure \ref{fig:network}. The main test environment was used for
the experiments. The retest environment was used to reconfirm critical results. Some
experiments showed unexpected results. Therefore, another testing machine was needed
to perform repeat testing at a strictly reduced complexity.

Figure \ref{fig:network} shows two test environments. The main test environment was used for the tests. If necessary, retests were done in a small, isolated retest environment to validate results.

\section{Results and comparative analysis}

\subsection{\textbf{Is the key in memory}} Davies analysed the encryption of 3 Windows ransomware samples and all samples used RSA+AES based encryption \cite{davies2020}. 
In contrast, this project showed that different algorithms were used in all Linux
ransomware samples. Only Icefire used AES in combination
with RSA. Cl0p used a hardcoded
RC4 master-key and Blackbasta used ChaCha20. A proposal for the determination of
ChaCha20 key material in memory dumps was found \cite{McLaren_2019}, but
reliable implementations are pending. The results were surprising and not expected.

\begin{figure}[htp]
    \centering
    \includegraphics[width=8.84cm]{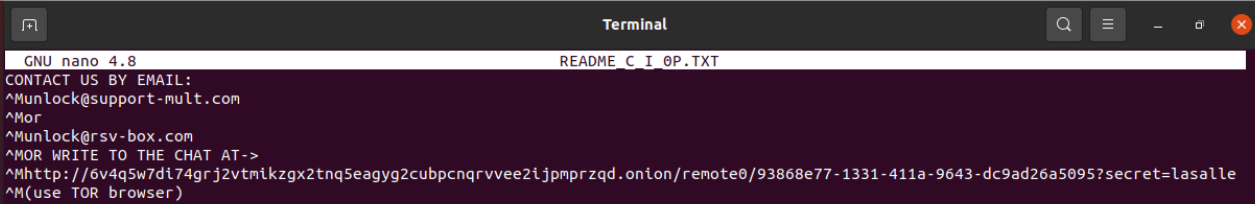}
    \caption{Ransom Message of Cl0p Linux ransomware}
    \label{fig:clopransom}
\end{figure}

All ransomware samples did not display visual ransom messages, similar to the Cl0p ransom message as seen in Figure \ref{fig:clopransom}.

A list of the identified cipher systems and the key extraction results is displayed in Table \ref{keymanagement}. 
\begin{center}
\begin{tabular}{ l l l l l}
 \textbf{Sample} & \textbf{Cipher}  &  \textbf{Keys extracted}\\
   \hline
 Icefire       & AES    & 11 \\  
 Cl0p             & RC4     & 0 \\ 
 Blackbasta   & ChaCha20    & 0
\end{tabular}
\captionof{table}{Key Management Systems of Linux ransomware}\label{keymanagement}
\end{center}

All ransomware samples created for each file new encryption keys. 
During the experiment Icefire was observed to clean the memory immediately after finishing encryption of one file.
Dynamic forensic investigation turned out to be very ineffective and effortful and was only successful partially for Icefire ransomware sample.

\subsection{\textbf{How long is the key present}} Windows-based ransomware tended to use same key for encryption and usually cleaned up the RAM at a late point in time \cite{davies2020}. The key disappeared from RAM only after a restart or after the encryption was completely finished. Thus chances to obtain the encryption key during dynamic forensic investigation and Live monitoring of RAM were realistic. In comparison, the Linux ransomware generated individual keys and used various encryption methods. Dynamic forensic methods were successful only for Icefire ransomware sample. The Icefire Linux ransomware maintained a clean memory, the encryption key was present during the encryption and disappeared immediately after. Tracking of keys for Cl0p and Blackbasta ransomware samples was not successful. An overview of the Live forensic monitoring results is 
shown in Table \ref{durationpresence}.
\begin{center}
\begin{tabular}{ l l l}
 \textbf{Sample} & \textbf{Becomes present}  &  \textbf{Duration of presence}\\
   \hline
 Icefire       & 5 seconds    & during file encryption \\  
 Cl0p             & unknown     & unknown \\ 
 Blackbasta   & unknown    & unknown
\end{tabular}
\captionof{table}{Duration of keys in RAM}\label{durationpresence}
\end{center}

Overall, Live (dynamic) forensic techniques turned out to be ineffective and inefficient for Linux ransomware.
Static forensic methods were more successful to obtain the key and to identify implementation errors in the ransomware sample. 

\subsection{\textbf{Does the key decrypt the files}} 
The Linux ransomware had weaknesses in key management. Mistakes in the implementation were found for 2 Linux ransomware variants and made decryption possible.
The decryption of Blackbasta \cite{lamsouber} and Cl0p ransomware files was successful \cite{sentinelone}. Often, it is no longer of any use that the data can be decrypted for free months or years later. Victims must immediately ensure continued operations. Ransomware attacks are therefore successful as long as there is an operational outage that forces the victim to pay at that moment. An overview of the decryption results is listed in Table \ref{decrpytionresults}.

\begin{center}
\begin{tabular}{ l l l}
 \textbf{Sample} & \textbf{Decrypted}  &  \textbf{Reason}\\
   \hline
 Icefire       & no    & RSA public key used \\  
 Cl0p             & decrypted     & hardcoded RC4 master-key \\ 
 Blackbasta   & decrypted    & decryptor leaked
\end{tabular}
\captionof{table}{Decryption results}\label{decrpytionresults}
\end{center}

\subsection{\textbf{Does the sample spread throughout the network.}} 
The research showed that all Linux ransomware samples did not attempt to spread
across the network. To summarise, the ransomware samples were primarily designed to perform the
encryption step for a specific victim. Delivery, lateral movement and feedback of
successful encryption therefore occur outside of ransomware capabilities.
The results were also compared with the results for Windows Active Directory domain
services \cite{McDonald_2022}. Windows ransomware was
capable to detect attached file shares and to encrypt the data inside. Linux ransomware
was not capable of doing this. WannaCry is known to contact command and control
servers (CC). If this step fails, the ransomware usually starts the encryption
immediately \cite{McDonald_2022}. Communication to CC servers is also
confirmed for TeslaCrypt \cite{skuratovich} and Jigsaw ransomware \cite{ashdown}.

An overview of the communication and network trace results is listed in Table \ref{networkspread}. 
\begin{center}
\begin{tabular}{ l l l l}
 \textbf{Communication detected?} & \textbf{Icefire}  &  \textbf{Cl0p} & \textbf{Blackbasta} \\
   \hline
 C\&C server        & no    & no & no \\  
 SSH                & no    & no & no \\ 
 FTP                & no    & no & no\\ 
 Samba              & no    & no & no \\ 
 WebServices        & no    & no & no \\ 
 Lateral Movement   & no    & no & no \\ 
 Key Exchange       & no    & no & no \\ 
 Cryptocurrency     & no    & no & no
\end{tabular}
\captionof{table}{Network communication \& lateral movement}\label{networkspread}
\end{center}

Windows ransomware is known to communicate to an command and control server for key management or to automatically scan the status of cryptocurrency wallets.
Such features were not observed for the 3 Linux ransomware samples. All samples did not communicate to external servers.

In conclusion, none of the ransomware samples communicted to external services. 
To validate the results, the tests were repeated in the isolated retest environment, allowing to communicate to the internet.

\subsection{\textbf{What is the impact of the encryption}}
The impact of Linux ransomware was compared with the research by McDonald et al., who analysed the
impact of Windows ransomware on Windows Active Directory Domain Services.
Windows ransomware had no impact on logon services. Users were
still able to log in. For Linux, the Cl0p ransomware encrypted the user files in such a way, that a GUI-based login was not possible anymore.

Windows ransomware was also able to move laterally to some extent,
to detect file shares and web server files (e.g. wwwroot of the IIS web server) and encrypt them as
well. The selected Linux ransomware samples were not able to do so. Both the Linux and Windows ransomware had no impact on the basic operation of the
web server itself.

Windows ransomware tends to restart after successful
encryption and to display a visual message. For Linux such behaviour was not
observed.

The impact on the Linux system was limited by the defined target of the Linux ransomware authors.
The Cl0p and Icefire ransomware encrypted files mainly in the user
directory, even when running with administrative privileges. Blackbasta encrypted only files in {\fontfamily{pcr}\scriptsize\selectfont /vmfs/volumes }, even if run as root.

All Linux ransomware samples did not utilise the full potential of the permissions being
executed. Even when running with full permissions,
important paths were not affected. The MySQL databases, SSH data, FTP data and
Samba shares were not affected. But especially in the enterprise sector, it is common
practice to connect external databases and storage and not to store any data under /home
or /root.

An overview of the impact on the system is displayed in Table  \ref{impact}. 
\begin{table*}
\begin{center}
\begin{tabular}{ p{0.2\textwidth} p{0.2\textwidth} p{0.2\textwidth} p{0.1\textwidth}}
 \textbf{Negative impact?} & \textbf{Icefire}  &  \textbf{Cl0p} & \textbf{Blackbasta} \\
   \hline
 Login                              & no    & GUI login fails & no \\  
 Network                            & no    & no & no \\ 
 Applications                       & Audio\&store  crashed  & Many apps crashed & no\\
 Encryption as user                 & {\fontfamily{pcr}\scriptsize\selectfont \$HOME }    & none, but permission  & none\\ 
 &       &   manipulated          &  \\  
 Encryption as root            & {\fontfamily{pcr}\scriptsize\selectfont \$HOME }     & {\fontfamily{pcr}\scriptsize\selectfont \$HOME } & {\fontfamily{pcr}\scriptsize\selectfont /vmfs/volumes }
\\
                                    &   {\fontfamily{pcr}\scriptsize\selectfont /usr }     &   {\fontfamily{pcr}\scriptsize\selectfont /root }         &  \\  
                                    &  {\fontfamily{pcr}\scriptsize\selectfont /root }     &   {\fontfamily{pcr}\scriptsize\selectfont /snap }          &  \\

 WebServer files                    & no    & no & no \\ 
 MySQL files                        & no    & no & no \\ 
 Automatic reboot?                  & no    & no & no
\end{tabular}
\captionof{table}{Decryption results}\label{impact}
\end{center}
\end{table*}

\section{Impact for IoT systems}
The results achieved for Linux ransomware serves an execellent basic understanding and allows to draw implications for the IoT industry \cite{gerodimos2023iot}. Security reasearches have recently executed ransomware on Bosch IoT devices directly, which was based on NEXO-OS Linux distribution \cite{boschrexrothhack}. This research demonstrates, that ransomware on IoT devices directly is achievable.

\subsection{\textbf{IoT threat overview}} Based on the findings in this research paper, it can be assumed that cybercriminals are more likely to simply block operations for a certain, unbearable period of time until the victim is willing to pay the ransom. While Windows ransomware is developed generically and can be widely used, the situation is different for Linux. Linux is individual in details. Therefore, Linux ransomware must be individually tailored and developed for the respective attack target. Attacks are only worthwhile if they promise considerable financial gain for the hackers. It can be assumed that attacks on IoT solutions are worthwhile if it is a "low hanging fruit" due to many vulnerabilities, or if the victim has reached such a large scale that even complex operations promise considerable profits. In any case, the barrier and effort for hackers is significantly higher than in comparison to Windows, which can sometimes be assembled using a modular principle.

A lower threat is seen for IoT solutions that have good basic security and are only averagely visible on the market. It can be assumed that costs and profit opportunities are not present for the attackers. Medium-sized manufacturers that meet basic IT security requirements are particularly at an advantage here.

\subsection{\textbf{IoT attack vectors}} An IoT solution can be attacked in 3 different ways: The IoT end device directly, the IoT gateway or the IoT cloud infrastructure. Attacks on the end device are often not scaleable and require physical or direct access, if the end device is not directly exposed to the internet. 

Attacks on IoT gateways also often lack scalability, but could be possible if attackers gain access to the victim's network through other means. Getting access to the victim's network is effortful, and the victim may not have big willingness to pay the ransom. The other option could be to just throw-away a low-cost device.
The third attack path would target the cloud infrastructure with the highest likelyhood for an scalable attack. In this case, the below-mentioned recommendations in Section \ref{sec:5} are directly applicable.

\section{Discussion and Conclusion} \label{sec:5}
In this paper, we outlined major differences between Windows and Linux ransomware.
The results enables IT responsibles to evaluate the risk in their area and take appropriate action.

We diversified the encryption methods. The experiments demonstrated diverse
key management and encryption methods for Linux systems. Only Icefire ransomware is confirmed to use AES encryption. Cl0p used a hardcoded RC4 symmetric master-key and Blackbasta
used ChaCha20 in combination with RSA. The results contradicted those of other
researchers compared to Windows operating systems.
As a result, the different use of encryption methods poses major practical difficulties for
live forensic investigations. Previously proven methods and tools that were mainly used for
RSA and AES can no longer be used. Suitable, ready-to-use implementations for
obtaining RC4 or Chacha20 keys from memory dumps were not found during the
literature search. 

Linux ransomware
development appears to be in its early stages and is expected
to progress and reach a similar level of maturity to Windows-based malware. The samples examined always serve a predefined goal and do
not exploit the full potential for damage. The following recommendations can be made to protect Linux systems.

\textbf{Avoid HOME directories.} To reduce the risk of a ransomware attack, it is
recommended not to store important data under /home/user or /root.

\textbf{Separate and restrict permissions and data access.} It also became apparent that the
ransomware was unable to independently take over other processes or user rights. 
Each application should have separate storage space and run with different users.

\textbf{ Avoid using privileged users.} Cl0p and Blackbasta could only encrypt the data if they
were run with admin rights. When the Icefire ransomware was run as a normal user, the
attack was only limited to the user's specific home directory.

\textbf{ Focus on identifying backdoors.} All 3 ransomware variants were unable to spread
independently across the network. Other computers on the network were unaffected.
This suggests that cybercriminals are targeting the victims and may have already
created backdoors to maintain access. Therefore, forensic resources should be focused
on identifying such backdoors.

\textbf{ Shut down first.} Weighing up the chances of forensic investigations and the risks, it is advisable to shut
down the infected Linux machine and surrounding systems as quickly as possible and
not to act hesitantly. Shutting down the system increases the chances of obtaining the
ELF ransomware binary for static analysis. Static analysis in particular proved to be
helpful in the experiments.
The chances that the key can be obtained during forensic investigations of the RAM are
slim.

IoT gateways should separate the operating system in a read-only partition. Installation and execution should be allowed only for signed software artifacts. The public key infrastructure and signing procedures should be operated separately. An offline public key infrastructure for signing the firmware and software is recommended. Regular updates of the IoT cloud infrastructure are required to maintain low exposure to vulnerabilities and reduce risk of lateral movement.

Security hygiene in development is essential to protect access to the IoT cloud infrastructure. Developers should install and use only approved and verified software and introduce secrets management solutions with secrets rotation. Development applications and tools should not be accessible throughout the internet.

\bibliographystyle{IEEEtran}
\bibliography{bibliography} 

\begin{thebibliography}{10}
\providecommand{\url}[1]{#1}
\csname url@samestyle\endcsname
\providecommand{\newblock}{\relax}
\providecommand{\bibinfo}[2]{#2}
\providecommand{\BIBentrySTDinterwordspacing}{\spaceskip=0pt\relax}
\providecommand{\BIBentryALTinterwordstretchfactor}{4}
\providecommand{\BIBentryALTinterwordspacing}{\spaceskip=\fontdimen2\font plus
\BIBentryALTinterwordstretchfactor\fontdimen3\font minus \fontdimen4\font\relax}
\providecommand{\BIBforeignlanguage}[2]{{%
\expandafter\ifx\csname l@#1\endcsname\relax
\typeout{** WARNING: IEEEtran.bst: No hyphenation pattern has been}%
\typeout{** loaded for the language `#1'. Using the pattern for}%
\typeout{** the default language instead.}%
\else
\language=\csname l@#1\endcsname
\fi
#2}}
\providecommand{\BIBdecl}{\relax}
\BIBdecl

\bibitem{lella2023enisa}
I.~Lella, C.~Ciobanu, E.~Tsekmezoglou, M.~Theocharidou, E.~Magonara, A.~Malatras, R.~Svetozarov~Naydenov \emph{et~al.}, ``Enisa threat landscape 2023: July 2022 to june 2023,'' 2023.

\bibitem{ryan2021ransomware}
M.~Ryan, \emph{Ransomware Revolution: The Rise of a Prodigious Cyber Threat}.\hskip 1em plus 0.5em minus 0.4em\relax Springer, 2021.

\bibitem{chainalaysis}
\BIBentryALTinterwordspacing
Chainalysis, ``Ransomware revenue down as more victims refuse to pay,'' 2023. [Online]. Available: \url{https://www.chainalysis.com/blog/crypto- ransomware-revenue- down-as-victims-refuse-to-pay/}
\BIBentrySTDinterwordspacing

\bibitem{thewhitehouse}
\BIBentryALTinterwordspacing
T.~W. House, ``International counter ransomware initiative 2023 joint statement,'' 2023. [Online]. Available: \url{https://www.whitehouse.gov/briefing-room/ statements-releases/2023/11/01/international-counter-ransomware- initiative-2023-joint-statement/}
\BIBentrySTDinterwordspacing

\bibitem{europol}
Europol, ``Cyber-attacks: the apex of crime-as-a-service, europol spotlight report series,'' 2023.

\bibitem{glenny}
\BIBentryALTinterwordspacing
M.~Glenny, ``The untold history of today’s russian-speaking hackers.'' \emph{Financial Times}, 2023. [Online]. Available: \url{https://www.ft.com/content/9ac188be-8bcf-4b5a-8051- 10563683b979}
\BIBentrySTDinterwordspacing

\bibitem{statistamarketshare}
\BIBentryALTinterwordspacing
Statista, ``Marketshare of leading operating systems worldwide from january 2009 till july 2023.'' 2023. [Online]. Available: \url{https://de.statista.com/statistik/daten/studie/157902/umfrage/ marktanteil-der- genutzten- betriebssysteme-weltweit-seit-2009/}
\BIBentrySTDinterwordspacing

\bibitem{thaeler2023enhancing}
A.~Thaeler, Y.~Yigit, L.~A. Maglaras, B.~Buchanan, N.~Moradpoor, and G.~Russell, ``Enhancing mac os malware detection through machine learning and mach-o file analysis.''\hskip 1em plus 0.5em minus 0.4em\relax Institute of Electrical and Electronics Engineers, 2023.

\bibitem{statistamarketshareserver}
\BIBentryALTinterwordspacing
Statista, ``Share of the global server market by operating system in 2018 and 2019,'' 2019. [Online]. Available: \url{Retrieved from https://www.statista.com/statistics/915085/global- server- share-by-os/}
\BIBentrySTDinterwordspacing

\bibitem{fortunebusiness}
\BIBentryALTinterwordspacing
F.~B. Insights, ``Server operating system market volume, share \& covid-19 impact analysis, by operating system (windows, linux, unix, and others), by virtualization status (virtual machine, physical, and virtualized), by subscription model (non-paid subscription and paid subscription), by enterprise type (large enterprises and small \& medium enterprises), and regional forecast, 2023-2030,'' 2023. [Online]. Available: \url{https://www.fortunebusinessinsights.com/serveroperating-system-market- 106601}
\BIBentrySTDinterwordspacing

\bibitem{terefos}
\BIBentryALTinterwordspacing
A.~Terefos, ``Cl0p ransomware targets linux systems with flawed encryption | decryptor available.'' 2023. [Online]. Available: \url{https://www.sentinelone.com/labs/cl0p-ransomware-targets-linux-systems-with-flawed-encryption-decryptor-available/}
\BIBentrySTDinterwordspacing

\bibitem{gdata}
\BIBentryALTinterwordspacing
G.~D. Software, ``G data threat report: Significant increase in linux ransomware.'' G Data Software, Tech. Rep. [Online]. Available: \url{https://www.gdatasoftware.co.uk/news/2022/08/37568-g-data-threat-report-significant-increase-in-linux-ransomware}
\BIBentrySTDinterwordspacing

\bibitem{trendmicro}
\BIBentryALTinterwordspacing
T.~Micro, ``Defending the expanding attack surface. trend micro 2022 midyear cybersecurity report.'' Tech. Rep., 2022. [Online]. Available: \url{https://documents.trendmicro.com/assets/rpt/rpt-defending-the-expanding- attack-surface-trend-micro-2022-midyear-cybersecurity-report.pdf}
\BIBentrySTDinterwordspacing

\bibitem{johansen}
G.~Johansen, \emph{Digital Forensics and Incident Response: Incident Response Tools and Techniques for Effective Cyber Threat Response.}, 3rd~ed.\hskip 1em plus 0.5em minus 0.4em\relax Birmingham: Packt Publishing, 2022.

\bibitem{internalbsipaper}
\BIBentryALTinterwordspacing
M.~Roth, ``Internal paper of the bsi. final report: Anhalt- bitterfeld paralyzed for too long by cyber attack.'' \emph{MITTELDEUTSCHER RUNDFUNK}, 2023. [Online]. Available: \url{https://www.mdr.de/nachrichten/sachsen-anhalt/dessau/podcast-cyberkatastrophe-bsi-kritik-landkreis-100.html}
\BIBentrySTDinterwordspacing

\bibitem{davies2020}
S.~R. Davies, ``Evaluation of live forensic techniques in ransomware attack mitigation,'' 2020.

\bibitem{delamotte}
\BIBentryALTinterwordspacing
A.~Delamotte, ``Icefire ransomware returns | now targeting linux enterprise networks,'' \emph{Sentinel LABS}, 2023. [Online]. Available: \url{https://www.sentinelone.com/labs/icefire-ransomware-returns-now-targeting-linux-enterprise-networks/}
\BIBentrySTDinterwordspacing

\bibitem{umawing}
J.~Umawing, ``Blackbasta is the latest ransomware to target esxi virtual machines on linux,'' \emph{MalwareBytes}, 2022.

\bibitem{McLaren_2019}
\BIBentryALTinterwordspacing
P.~McLaren, W.~J. Buchanan, G.~Russell, and Z.~Tan, ``Deriving chacha20 key streams from targeted memory analysis,'' \emph{Journal of Information Security and Applications}, vol.~48, p. 102372, Oct. 2019. [Online]. Available: \url{http://dx.doi.org/10.1016/j.jisa.2019.102372}
\BIBentrySTDinterwordspacing

\bibitem{lamsouber}
L.~Lamsouber, ``What is a ransomware and how does it work?'' 2023.

\bibitem{sentinelone}
\BIBentryALTinterwordspacing
S.~One, ``Cl0p-elf-decryptor,'' 2023. [Online]. Available: \url{https://github.com/SentineLabs/Cl0p-ELF-dcryptor/blob/main/clop\_linux\_file\_decr.py}
\BIBentrySTDinterwordspacing

\bibitem{McDonald_2022}
\BIBentryALTinterwordspacing
G.~McDonald, P.~Papadopoulos, N.~Pitropakis, J.~Ahmad, and W.~J. Buchanan, ``Ransomware: Analysing the impact on windows active directory domain services,'' \emph{Sensors}, vol.~22, no.~3, p. 953, Jan. 2022. [Online]. Available: \url{http://dx.doi.org/10.3390/s22030953}
\BIBentrySTDinterwordspacing

\bibitem{skuratovich}
S.~Skuratovich, ``Check point looking into teslacrypt v3.0.1,'' \emph{CHECK POINT}, 2016.

\bibitem{ashdown}
\BIBentryALTinterwordspacing
D.~Ashdown, ``Jigsaw ransomware analyses,'' 2021. [Online]. Available: \url{https://www.cyberdonald.com/post/jigsaw-ransomware-analyses}
\BIBentrySTDinterwordspacing

\bibitem{gerodimos2023iot}
A.~Gerodimos, L.~Maglaras, M.~A. Ferrag, N.~Ayres, and I.~Kantzavelou, ``Iot: Communication protocols and security threats,'' \emph{Internet of Things and Cyber-Physical Systems}, 2023.

\bibitem{boschrexrothhack}
\BIBentryALTinterwordspacing
N.~N. Labs, ``Vulnerabilities on bosch rexroth nutrunners may be abused to stop production lines, tamper with safety-critical tightenings,'' 2024. [Online]. Available: \url{https://www.nozominetworks.com/blog/vulnerabilities-on-bosch-rexroth-nutrunners}
\BIBentrySTDinterwordspacing

\end{thebibliography}
\end{document}